**Implications of adopting plane angle as a base quantity in the SI**

Paul Quincey and Richard J. C. Brown

Environment Division, National Physical Laboratory, Hampton Road, Teddington, TW11 0LW, United Kingdom.

Tel: +44 (0)20 8943 6788; E-mail address paul.quincey@npl.co.uk

**Abstract**

The treatment of angles within the SI is anomalous compared with other quantities, and there is a case for removing this anomaly by declaring plane angle to be an additional base quantity within the system. It is shown that this could bring several benefits in terms of treating angle on an equal basis with other metrics, removing potentially harmful ambiguities, and bringing SI units more in line with concepts in basic physics, but at the expense of significant upheaval to familiar equations within mathematics and physics. This paper sets out the most important of these changes so that an alternative unit system containing angle as a base quantity can be seen in the round, irrespective of whether it is ever widely adopted. The alternative formulas and units can be treated as the underlying, more general equations of mathematical physics, independent of the units used for angle, which are conventionally simplified by implicitly assuming that the unit used for angle is the radian.

**Introduction**

<u>Base quantities</u>

The SI currently considers seven quantities to be base quantities - time, length, mass, electric current, temperature, amount of substance, and luminous intensity (BIPM, 2006). This situation will not change in the process of redefining the units for mass, electric current, temperature and amount of substance, which is expected to be completed over the next few years (BIPM, 2016). The general principle is that each base quantity has a clearly defined measurement unit, referred to as a base unit, and all other quantities are measured using units that are derived from a combination of these base units.

The set of seven is not the only possible set of base quantities for a system of units, and indeed the SI has increased the set it uses since it was established in 1960, with the addition of amount of substance in 1971. The choice of base quantities, both the quantities themselves and the number of them in the set, can be seen as a compromise between several factors:

- Limiting the number of precise unit definitions required, bearing in mind that the system of units should be capable of underpinning the most accurate measurements achievable in the scientific and technological worlds, and that the definitions must be capable of being realised with high precision within metrology laboratories;

- Reflecting the quantities that are most fundamental and independent as perceived within the laws of nature (especially physics);

- Retaining clarity for quantities that are best considered distinct, even though they are not independent, for example length and time, which can be seen as related quantities in the space-time of special relativity;

- Allowing dimensional analysis, described in more detail below. This valuable tool relies on all quantities being described in terms of a small number of dimensions. The tool is most valuable when these dimensions correspond to the base quantities of the system of units being used. Reducing the number of base quantities can reduce the power of dimensional analysis;

- Satisfying user communities. A system of units intended for global use must be acceptable to as wide a range of user communities as is practical. Decisions on whether or not to consider certain quantities as base quantities, with their consequences for how "fundamental" a quantity is viewed as being, and for the scientific equations that are used (as described in the next section), must take this into account.

Some history relating to systems of units: dimensional analysis and the case of electromagnetism

Measurement units have been around since antiquity, but systems of units have only been proposed relatively recently. We use the term "system of units" in the sense of a rational system that meets the practical needs always met by measurement units, but in an efficient way that makes use of the known relationships between different quantities. As a simple example, the basic unit for volume in a system of units will be the (base unit for length)$^3$, unlike an *ad hoc* set of units when lengths might be measured in inches, and volumes in gallons, with a conversion factor between gallons and cubic inches fixed by convention.

The number of base units within a system of units can be seen as closely linked to the concept of dimensions. The basic aspects of dimensional analysis are set out in Section 1.3 of the current SI Brochure (BIPM, 2006). In essence, dimensional analysis can perform two functions. Firstly, and more prosaically, by treating all quantities as being composed of a small number of independent dimensions, such as mass (**M**), length (**L**) and time (**T**), it is relatively simple to check whether an equation contains a set of terms that make the dimensions balance on both sides. Secondly, the requirement that the laws of nature be written in equations where the types of quantity on either side must match - the "great principle of similitude", as Lord Rayleigh put it (Rayleigh, 1915) - can provide deep insights into the form these laws must take. To take an example of this latter function from Rayleigh's paper, the form of the law governing light scattering by small particles, now known as Rayleigh scattering, which is that the intensity of scattered light is inversely proportional to the fourth power of the wavelength, can be deduced simply by considering the dimensions involved. Although generally taught as a peripheral part of science courses, dimensional analysis has had a long and distinguished role in science, from early presentations (e.g. Rayleigh, 1915; Bridgman, 1922) to more recent advocacy (e.g. Robinett, 2015).

The most significant development in the history of systems of units has been the incorporation of electromagnetism into the older mechanical system of mass, length and time, a topic that still causes some controversy (Jackson, 1999). The suggestion that no new dimensions, and therefore no new base quantities, were required to do this was made by Gauss in around 1832. The SI instead includes a new base quantity, electric current, in part because it is clear that electric charge is a fundamentally distinct property of matter that cannot reasonably be described as a combination of mass, length and time.

We can see from the example of electromagnetism the general consequences of adding a base quantity to a system of units. To start from the law describing the force between two electric charges:

$F \propto Q_1 Q_2 / d^2$             Equation 1

where $F$ is the force (dimension **MLT$^{-2}$**), $Q_1$ and $Q_2$ are the electric charges, $d$ is the distance between them (dimension **L**) and α denotes proportionality.

If we follow the Gaussian approach, we can assign to $Q$ the dimension **M$^{0.5}$L$^{1.5}$T$^{-1}$** to make the equation dimensionally balanced. The relationship then just needs a dimensionless constant to turn it into an equation, and this constant can be given the value of exactly one by the suitable choice of the unit for $Q$.

By choosing to give an electrical quantity its own dimension (electric current in the case of the SI), it is no longer possible to make Equation 1 dimensionally balanced simply by adding a dimensionless constant. With $Q$ having dimensions **IT**, where **I** represents current, the equation must have the form:

$F = k_1 Q_1 Q_2 / d^2$                Equation 2,

where $k_1$ is a dimensional constant with the dimensions **ML$^3$T$^{-4}$I$^{-2}$**. Within the SI this constant is the familiar $1/4\pi\varepsilon_0$.

The key point here is that even though the underlying physical relationship between the quantities is the same within the two systems, when an extra dimension is introduced in the system of units, a dimensional constant must also be added to make the equation describing the relationship dimensionally balanced. This will be a crucial point when considering the addition of angle[1] as an extra dimension. Equations must change when a new dimension is introduced; it is not as simple as just elevating a unit to a higher status and leaving everything else unchanged.

The lack of widespread changes to scientific equations when amount of substance was declared a base quantity in 1971 might seem to contradict this statement. However, the effect of this change was to bring the equations of chemistry into the same system as those for physics, leaving the equations of physics and chemistry themselves relatively unaffected. It had the benefit of introducing dimensional analysis to chemistry and providing a formal distinction between intensive and extensive quantities (such as 'molar mass' and the 'mass of a mole') that previously could have been confused (Brown and Brewer, 2015). Angle is far more embedded in the equations of mathematical physics than amount of substance, and the changes would be far more widespread.

---

[1] The term *angle* in this paper is used exclusively to mean plane angle, not solid angle.

**The case for making angle a base quantity**

As described in an earlier paper (Quincey, 2016), angle is currently treated within the SI as a dimensionless quantity, seen either as a quantity whose value is "counted" in radians (which are not necessarily whole radians), or as the ratio of arc length to radius for a circle centred on the angle being determined. In practice, angle is something that is measured using an instrument with a scale, which is usually marked in degrees, rather than counted in radians or calculated as a ratio. From this standpoint, angle is eminently suited to being treated like other dimensional quantities such as length and time. As a measurable quantity, its status as dimensionless is anomalous, and the reasons for this can be seen in its origins in mathematics rather than experimental science. The situation leads to genuine complications concerning when the unit symbol rad should be used or replaced by the number one, and whether the unit Hertz refers to some kind of cycle or revolution per second, or to a radian per second.

Given also that the radian cannot be definitely derived from other units – the equation 1 rad = 1 m/m does not uniquely define the radian (Quincey, 2016) - one solution to these practical complications would be to remove the anomaly and declare angle to be a base quantity, with an associated base unit: the radian.

Perhaps equally persuasive is the fact that, because angle is treated as dimensionless, the units for physical quantities like torque and angular momentum give no indication of their angular quality, which if explicit would bring greater clarity to the underlying physics. For example, within the current SI, *energy* and *torque* have the same units. By making angle a base quantity, it is possible instead to make the unit for *energy* the same as for *torque* x *angle*, analogous to *energy* being *force* x *distance*.

Similarly, instead of the quantities *action* and *angular momentum* having the same units, as is the case within the current SI, *angular momentum* could become *action* per unit *angle*, in direct analogy to *linear momentum* being *action* per unit *length*, and *energy* being *action* per unit *time*. The change would consequently be beneficial in terms of dimensional analysis, in addition to providing a better description of some physical quantities.

The case for treating angle as a base quantity has been made many times over a long period, for example by Brinsmade (1936), Romain (1962), Eder (1982), Torrens (1986), Brownstein (1997), Foster (2010) and Mohr and Phillips (2015).

**The implications of making angle a base quantity.**

It has already been pointed out that the introduction of angle as a base quantity cannot be done without other changes also taking place, to make equations balance dimensionally. There are several ways of doing this.

<u>Changing the dimensions of familiar quantities</u>

One approach, set out in Eder (1982), is to leave the equations of mathematics and physics unchanged, and to adjust the dimensions of other quantities accordingly. The most striking example of this is that while lengths in general retain the dimension *length*, radii would acquire the dimensions of *length/angle*.

A simple example of the difficulties raised by this approach is shown in Figure 1. An isosceles triangle is placed within a circle so that its equal sides form radii of the circle. The base of the triangle is now the chord c, and there is an associated length of arc denoted by s. What dimension should be used for the lengths marked b? If only the triangle is considered, they are surely simply *length*; if the arc is being considered, we are being asked to treat b as having the dimension *length/angle*. We are therefore being asked to consider that b should be measured either in metres or in metres per radian, depending on the circumstances.

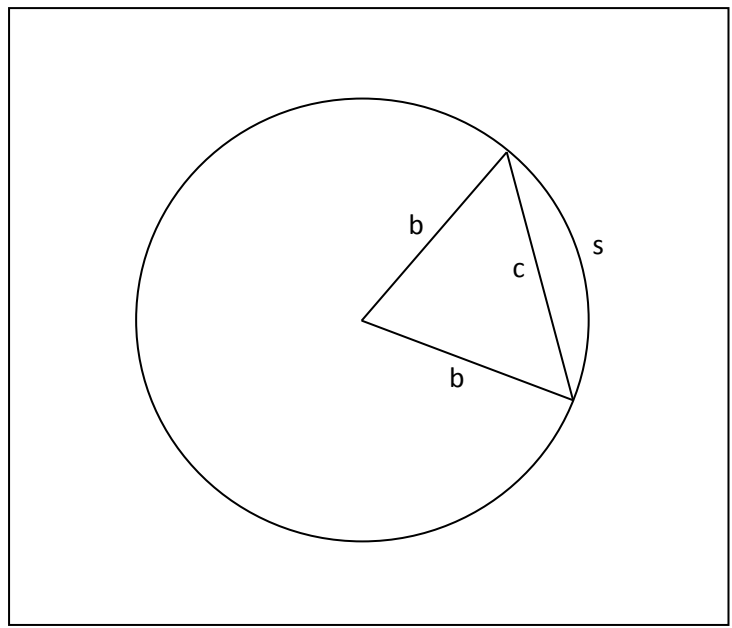

Figure 1: Arc and chord diagram

Although this system can be made to work more or less correctly, *ad hoc* rules are needed to assign dimensions to various quantities, in a way that is not necessary within the current SI. The result of this is to effectively invalidate the technique of dimensional analysis, which is a very high price to pay, and also to make the underlying physics less clear. Returning to the example of the force between electrical charges, it would be like making Equation 2 balance by declaring that, in this situation, the separation between charges should be considered to have the dimensions **M**$^{-0.5}$**L**$^{-0.5}$ **T**$^2$**I**, instead of simply **L**.

A second approach is available, which retains the validity of dimensional analysis by the use of a dimensional constant, and which is also far more transparent. Only this second approach is considered further here.

Introducing a dimensional constant

The second approach is to adjust for the change in dimensions by adding a dimensional constant to the equations of mathematics and physics that are valid within the current SI. Following Torrens' notation (1986), the $\eta$-factor (equal to 1 rad$^{-1}$, with the dimension **A**$^{-1}$, where **A** is the dimension of plane angle) will need to be included in many formulas and equations[2]. It will take the numerical value 1 when the relevant angle is expressed in radians, or π/180 when the angle is in degrees, or 2π when the angle is in revolutions. In effect, this would be the new defining constant used to incorporate angle as a base quantity within the SI. Unlike the other defining constants, its value would be set by mathematics rather than by precedent and experiment.

Self-consistent changes to the equations of mathematics and physics can be made in several different ways. One of these ways is set out here.

Mathematical functions.

The change within the SI would have no effect on purely mathematical functions such as

$$sin\, x = x - \frac{x^3}{3!} + \frac{x^5}{5!} - \ldots \quad \text{Equation 3}$$

However, to be compatible with the SI, when the quantity x is explicitly an angle θ (which may or may not be expressed in radians), the function would need to change to:

$$sin\, \theta = \eta\theta - \frac{\eta^3\, \theta^3}{3!} + \frac{\eta^5\, \theta^5}{5!} - \ldots \quad \text{Equation 4.}$$

Other trigonometric functions would change accordingly.

Differentials and integrals of these functions would also change, so that, for example,

$$\frac{d(sin\, \theta)}{d\theta} = \eta.\, cos\, \theta \quad \text{Equation 5}$$

Mathematical physics

Some of the changes required within mathematical physics are set out in Table 1. In most cases, these changes follow directly from the priority of matching each quantity to its most suitable dimensions. The main exception is a choice between whether to include an $\eta^2$ factor in the definition of moment of inertia, or to include it instead in the expressions that contain the moment of inertia. The latter option has been taken here, taking the view that moment of inertia is a property of an extended object about a specified axis, which does not intrinsically rely on a choice of angle unit.

---

[2] Some authors, starting with Brinsmade (1936), use the unit symbol rad as a dimensional constant within equations, with the value 1 rad. As no other unit symbol is used in this way, we feel that this is confusing, and that the explicit $\eta$-factor, which is the same as the □-factor in Brownstein (1997), is preferable.

| Quantity | Symbol | Current formula | New formula | Current SI unit | New SI unit |
|---|---|---|---|---|---|
| angle | $\vartheta$ | | | rad or 1 (one) | rad |
| angular velocity | $\omega$ | $= d\vartheta/dt$ | unchanged | rad.s$^{-1}$ or s$^{-1}$ | rad.s$^{-1}$ |
| arc length | s | $= r.\vartheta$ | $= \eta.r.\vartheta$ | m | unchanged |
| sector area | A | $= ½ r^2.\vartheta$ | $= ½ \eta.r^2.\vartheta$ | m$^2$ | unchanged |
| volume element in spherical coordinates | dV | $= r^2.\sin\varphi.dr.d\varphi.d\vartheta$ | $= \eta^2.r^2.\sin\varphi.dr.d\varphi.d\vartheta$ | m$^3$ | unchanged |
| tangential velocity | v | $= r.\omega$ | $= \eta.r.\omega$ | m.s$^{-1}$ | unchanged |
| centripetal acceleration | a | $= r.\omega^2$ | $= \eta^2.r.\omega^2$ | m.s$^{-2}$ | unchanged |
| moment of inertia | I | $= \Sigma m_i.r_i^2$ | unchanged | kg.m$^2$ | unchanged |
| angular momentum | **L** | $= $ **r** $\wedge$ **p** | $= \eta.$ **r** $\wedge$ **p** | kg.m$^2$.s$^{-1}$ | kg.m$^2$.s$^{-1}$.rad$^{-1}$ |
| | | $= I.\omega$ | $= \eta^2.I.\omega$ | kg.m$^2$.s$^{-1}$ | kg.m$^2$.s$^{-1}$.rad$^{-1}$ |
| torque | $\tau$ | $= $ **r** $\wedge$ **F** | $= \eta.$ **r** $\wedge$ **F** | kg.m$^2$.s$^{-2}$ | kg.m$^2$.s$^{-2}$.rad$^{-1}$ |
| | | $= I.d\omega/dt$ | $= \eta^2.I.d\omega/dt$ | kg.m$^2$.s$^{-2}$ | kg.m$^2$.s$^{-2}$.rad$^{-1}$ |
| rotational kinetic energy | | $= ½ I.\omega^2$ | $= ½ \eta^2.I.\omega^2$ | kg.m$^2$.s$^{-2}$ | unchanged |
| For comparison: | | | | | |
| linear momentum | **p** | | | kg.m.s$^{-1}$ | unchanged |
| force | **F** | | | kg.m.s$^{-2}$ | unchanged |
| action | S | | | kg.m$^2$.s$^{-1}$ | unchanged |
| energy | E | | | kg.m$^2$.s$^{-2}$ | unchanged |

Table 1: Proposed table of selected formulae and units within mathematical physics, if angle is adopted as a base quantity.

Note on the units for the Planck constant, $h$, and $\hbar$

Within the current SI system, both $h$ and $\hbar$ have the units kg.m$^2$.s$^{-1}$, or J.s. The question arises as to whether this would or should change if angle became a base quantity.

Simplistically, if we consider $E = h.f$ and $E = \hbar.\omega$ as the fundamental equations involving $h$ and $\hbar$, the change in status of angle would leave the units for $h$ unchanged, while, for $\hbar$, we would need to either assign $\hbar$ the units of angular momentum, kg.m$^2$.s$^{-1}$.rad$^{-1}$, or modify the equation to $E = \eta.\hbar.\omega$.

However, it has been suggested (e.g. Freeman, 1986) that, by analogy with angular frequency having the units rad/s, frequency should have the explicit unit cycle per second, abbreviated to cy/s, where cycle is treated as a new base quantity. It has alternatively been suggested that a cycle should be treated as a quantity of angle, equivalent to a revolution[3]. Starting from the equation $E = h.f$, these suggestions would mean either assigning $h$ the units kg.m$^2$.s$^{-1}$.cy$^{-1}$, or the units kg.m$^2$.s$^{-1}$.rad$^{-1}$, or changing the equation $E = h.f$ to $E = \eta.h.f$, for example.

In our view, the Planck constant $h$ is fundamentally a measure of action (Quincey, 2013), and should therefore have the units of action, kg.m$^2$.s$^{-1}$, or J.s, units that hold whether or not angle is considered a base quantity. This is supported by considering the de Broglie relation $p = h/\lambda$, where $p$ is linear momentum and $\lambda$ is wavelength. This equation can be considered to be as fundamental as the

---
[3] Ian Mills, private communication.

Planck equation, but does not involve the potentially confusing quantities of frequency or angular frequency. Here, $h$ also has the units $kg.m^2.s^{-1}$, or J.s. This provides an argument for continuing to treat the unit for frequency simply as $s^{-1}$, rather than cy/s, and at the same time retaining the equation $E = h.f$. It would also avoid the need to make changes to other familiar equations such as $v = f.\lambda$, relating the velocity, frequency and wavelength of a wave. Incidentally, the proposed redefinition of the kilogram in terms of the Planck constant also assumes that $h$ has units of $kg.m^2.s^{-1}$ or J.s.

The decision on the best units for $\hbar$ would be more of a free choice. There would be an argument for giving $\hbar$ the units of angular momentum, while $h$ retains those of action, so that the familiar equations $E = h.f$ and $E = \hbar.\omega$ can be retained, and the role of $\hbar$ as the natural unit of angular momentum is highlighted. $h$ and $\hbar$ would then be related by the equation $\hbar = \eta h/2\pi$.

**Conclusions**

It is entirely possible to include plane angle in the SI as an eighth base quantity, and this would bring several benefits in terms of treating angle on an equal basis with other metrics, removing potentially harmful ambiguities, and bringing SI units more in line with concepts in basic physics.

However, this would be at the expense of a major upheaval in terms of basic mathematical and physical equations, which is summarised in Table 1. We consider that it would be much harder to find global acceptance for these changes than for the changes to the SI base units currently being discussed. Moreover, incorporation of angle in this way would require an additional, rather strange, defining constant formulation of the radian as a base unit within the new SI, along the lines of "The radian, symbol rad, is the unit of angle; it is defined by taking the fixed numerical value of the ratio of the arc length of a circle to the product of its radius and the angle subtended at its centre, known as the η-constant, to be exactly 1, where the arc length and radius are expressed in the same units, and the angle is expressed in the unit rad".

As set out in the earlier paper (Quincey, 2016), we do not consider the upheaval to justify the benefits, at least not unless an extensive consultation outside the metrological community comes to the opposite conclusion. However, we consider it useful for this option to be set out clearly, both to prevent a decision to include angle as a base quantity within the SI being made without understanding the full implications, and to provide an alternative unit system that may give a useful new perspective to students and others who are interested in the relationship between scientific laws and systems of units. The suggested formulas and units in Table 1 can be treated as the underlying, more general equations of mathematical physics, independent of the units used for angle, which are conventionally simplified by implicitly assuming that the unit used for angle is the radian. These underlying equations could have specialised uses, for example within quantity-calculating software, where they would ensure that the software correctly handles units involving angle and frequency.

Acknowledgements

We would like to thank Ian Mills for stimulating discussions on this topic.